\documentclass[lettersize,journal]{IEEEtran}
\usepackage{amsmath,amsfonts}
\usepackage{algorithmic}
\usepackage{algorithm}
\usepackage{array}

\usepackage{textcomp}
\usepackage{stfloats}
\usepackage{url}
\usepackage{verbatim}
\usepackage{graphicx}
\usepackage{cite}
\usepackage{subfigure}
\usepackage{cleveref}
\usepackage[table,xcdraw]{xcolor}

\begin{document}

\title{SurfMyoAiR: A surface Electromyography based framework for Airwriting Recognition}
\author{Ayush Tripathi, Lalan Kumar, \IEEEmembership{Member, IEEE}, Prathosh A.P., and Suriya Prakash Muthukrishnan
\thanks{This work was supported in part by Prime Minister’s Research Fellowship (PMRF), Ministry of Education (MoE), Government of India with grant number PLN08/PMRF (to Ayush Tripathi).}
\thanks{This work involved human subjects or animals in its research. Approval
of all ethical and experimental procedures and protocols was granted by
the Institute Ethics Committee, All India Institute of Medical Sciences, New Delhi, India with reference number IEC-267/01.04.2022,RP-55/2022.}
\thanks{Ayush Tripathi is with the Department of Electrical Engineering, Indian Institute of Technology Delhi, New Delhi - 110016, India(e-mail: ayush.tripathi@ee.iitd.ac.in).}
\thanks{Lalan Kumar is with the Department of Electrical Engineering,
Bharti School of Telecommunication, and,
Yardi School of Artificial Intelligence, Indian Institute of Technology Delhi, New Delhi - 110016, India(e-mail: lkumar@ee.iitd.ac.in).}
\thanks{Prathosh A.P. is with the Department of Electrical Communication Engineering, Indian Institute of Science, Bengaluru - 560012, India(e-mail: prathosh@iisc.ac.in).}
\thanks{Suriya Prakash Muthukrishnan is with the Department of Physiology, All India Institute of Medical Sciences, New Delhi - 110016, India(e-mail: dr.suriyaprakash@aiims.edu).}}

\markboth{Submitted to IEEE Transactions on Instrumentation and Measurement}%
{}


\maketitle

\begin{abstract}
Airwriting Recognition is the task of identifying letters written in free space with finger movement. It is a dynamic gesture recognition with the vocabulary of gestures corresponding to letters in a given language. Electromyography (EMG) is a technique used to record electrical activity during muscle contraction and relaxation as a result of movement and is widely used for gesture recognition. Most of the current research in gesture recognition is focused on identifying static gestures. However, dynamic gestures are natural and user-friendly for being used as alternate input methods in Human-Computer Interaction applications. Airwriting recognition using EMG signals recorded from forearm muscles is therefore a viable solution. Since the user does not need to learn any new gestures and a large range of words can be formed by concatenating these letters, it is generalizable to a wider population. There has been limited work in recognition of airwriting using EMG signals and forms the core idea of the current work. The SurfMyoAiR dataset comprising of EMG signals recorded during writing English uppercase alphabets is constructed. Several different time-domain features to construct EMG envelope and two different time-frequency image representations: Short-Time Fourier Transform and Continuous Wavelet Transform were explored to form the input to a deep learning model for airwriting recognition. Several different deep learning architectures were exploited for this task. Additionally, the effect of various parameters such as signal length, window length and interpolation techniques on the recognition performance is comprehensively explored. The best-achieved accuracy was $78.50\%$ and $62.19\%$ in user-dependent and independent scenarios respectively by using Short-Time Fourier Transform in conjunction with a 2D Convolutional Neural Network based classifier. Airwriting has great potential as a user-friendly modality to be used as an alternate input method in Human-Computer Interaction applications.

\end{abstract}


\section{Introduction}

\subsection{Background}

The ability to communicate is one of the most important of all life skills that humans possess. With the rapid emergence and evolution of digital devices, interaction of humans with these devices has also increased. However, the medium of user input still remains traditional that includes keyboard, mouse, touchscreen. Hence, alternate input methods need to be explored for Human Computer Interaction (HCI) that can reduce the burden of carrying additional devices. This is to be achieved by modalities that can act as natural extensions of human cognition, thereby leading to seamless connection with the digital world\cite{MIT,FB}. The example includes virtual and augmented reality based devices where output of the system is directly fed to the user's eyes. This leaves no scope of using external peripherals for providing input to the system. Various ways have been explored for giving input to such devices. Providing speech-based input is one of the most commonly used solution\cite{SpeechVR}. However, in the presence of noise and reverberations, or in the case when the user suffers from a speech disorder, the performance of speech recognition systems degrades significantly \cite{Moore2018}. Additionally, this method is not feasible when user is required to maintain privacy, such as at a public place. An alternate approach that is suitable for the task is gesture recognition \cite{yasen2019systematic,cheok2019review}. Although privacy and silent transmission of information are taken care of by using gesture recognition, it suffers from additional limitations. A fixed dictionary of gestures is utilized for communication, thus limiting the range of interaction. Additionally, the user is required to learn the set of gestures and memorize them. Airwriting recognition, a special case of gesture recognition, overcomes these shortcomings and can be captured using various physiological signals.  

In this work, Electromyography (EMG) based airwriting recognition is explored. Airwriting is the process of writing letters in free space using unrestricted finger movements \cite{amma2013airwriting,7322243,tripathi2022imair,tripathi2021sclair}. EMG is a physiological signal generated due to muscle contraction and relaxation during the movement\cite{gohel2020review}. EMG signal can be acquired in two ways: (a) intramuscular EMG, where electrodes are inserted into the skeletal muscle and (b) surface EMG, where the electrodes are placed on the skin above the muscle.  Surface EMG (sEMG) has been utilized widely because of its non-invasive and user-friendly nature. sEMG has been used for applications such as gesture recognition\cite{li2021gesture}, prosthetic control\cite{samuel2018pattern}, rehabilitation\cite{campanini2020surface}, and human machine interaction\cite{simao2019review}. In airwriting, the vocabulary of gestures corresponds to letters in a particular language. Since the user is not required to learn any new set of gestures and a wide range of words can be formed by concatenating letters, airwriting can provide the user with a wide range of interaction capabilities. This makes such a system easy to use and generalizable to a larger section of population.

\subsection{Related Work}

The recognition of hand gestures using sEMG signals has garnered wide attention for various HCI applications including robot control\cite{boru2022novel}, rehabilitation\cite{zhou2021toward}, sign language recognition\cite{tateno2020development}, and user authentication\cite{9745921}. The field of hand gesture recognition can be further subdivided into two categories: static gesture recognition and dynamic gesture recognition\cite{xu2017review}. In static gesture recognition, the prime focus is on identification of gestures formed by specific hand shapes without any temporal dimension. There have been various attempts at tackling this problem by using either handcrafted features along with traditional machine learning based approaches\cite{simao2019review} or deep learning based approaches\cite{li2021gesture}. In \cite{8558109}, the authors proposed a 9-class gesture classification using time domain features and linear discriminant analysis classifier from 3-channel sEMG signals. In \cite{8768831}, a hand-gesture classification system using time-domain features and a neural network architecture using sEMG signals from a Myo Armband was proposed. In addition to handcrafted features, deep learning methods have been applied for the task with reasonably high performance. Convolutional Neural Network (CNN) has been widely used to extract the spatial relationship present in the multi-dimensional sEMG signals. In \cite{7759384}, the authors proposed a CNN-based gesture classification scheme using frequency domain features for application in robotic arm. An attention-based CNN and recursive neural network (RNN) architecture with six different types of images as input for sEMG-based gesture recognition was proposed by the authors in \cite{hu2018novel}. The authors in \cite{9422807} proposed a few-shot training strategy in order to minimize the need for re-calibration and allow the user to retrain the model for additional gestures. In \cite{9275331}, the authors investigated the use of sEMG signals recorded from the wrist for classification of different single-finger, multi-finger and wrist gestures for HCI applications.  Various studies also propose gesture classification schemes using different image representations of the sEMG signals such as Short-Time Fourier Transform (STFT) \cite{8630679}, Continuous Wavelet Transform (CWT) \cite{oh2021classification,pancholi2019improved}, Emperical Mode Decoposition \cite{9299282}, raw sEMG images \cite{geng2016gesture}, sEMG muscle activation maps \cite{8911244}, and grayscale sEMG images\cite{du2017surface}.  

Dynamic gesture recognition deals with recognition of hand motion trajectory in space. It becomes crucial to comprehensively consider position, shape and trajectory of the movement simultaneously, making the task of dynamic gesture recognition particularly challenging. In \cite{9658997}, a recognition model based on CNN trained with time-frequency images was proposed for identifying $5$ dynamic hand movements. Yang et.al. \cite{yang2021dynamic} proposed a multi-stream residual network (MResLSTM) for sEMG based classification of six dynamic gestures involving the entire hand movement. The problem of recognition of handwritten alphabets may be considered as a specific case of dynamic gesture recognition. A dynamic time warping based method was proposed in \cite{5627246} for handwriting recognition, that was further improved in \cite{li2013improvements}. Linderman et.al. \cite{linderman2009recognition} proposed a template matching framework for recognition of handwriting from sEMG signals. In \cite{beltran2020multi}, the authors proposed a methodology based on sEMG signals to recognize multi-user free-style handwriting characters. A CNN-LSTM based framework for classification of 36 handwritten characters (A-Z, 0-9) using filtered EMG signals was proposed in their study. Unlike the traditional setting of writing, airwriting recognition aims to identify characters written in free space with wrist/finger movements. The absence of a support for the finger during the writing process and lack of visual and haptic feedback adds a level of challenge to this problem. There have been various attempts by using inertial sensors \cite{yanay2020air,tripathi2022imair,tripathi2021sclair} and computer vision based approaches\cite{kim2021writing,MUKHERJEE2019217} for tackling the airwriting recognition problem. However, to the best of the authors' knowledge, sEMG based airwriting recognition has not been explored hitherto. Inspired by the success of the deep learning based approaches in sEMG based gesture recognition, different types of sEMG envelopes and time-frequency representations to serve as input to a deep learning model are explored for the purpose of airwriting recognition.

\subsection{Objectives and Contributions}

Airwriting is a dynamic gesture characterized by continuous motion of the wrist to write a specific letter. Unlike static gesture recognition, there has been limited focus on the recognition of dynamic gestures. The focus in this work is on the identification of the dynamic airwriting gestures from sEMG signals. First, the SurfMyoAiR dataset is established, which comprises of sEMG signals recorded from $50$ subjects during the task of airwriting. To the best of the authors' knowledge, this is the first instance of a large-scale dataset for the sEMG based airwriting recognition task. Subsequently, different processing strategies and deep learning architectures for recognition of airwritten letters are explored. The specific contributions of this article are listed below.

\begin{itemize}
    \item A surface Electromyography based dataset (SurfMyoAiR) recorded from the forearm muscles of $50$ subjects while writing the English uppercase alphabets ($10$ repetitions) in air is created. 
    
    \item The performance of different time-domain and time-frequency domain based approaches in conjunction with deep learning based classification schemes for the task of airwriting recognition is analyzed. 
    
    \item The effects of varying different parameters such as interpolation techniques, signal duration and window size are comprehensively explored, and the optimum choice of parameters suited for the airwriting recognition task are reported. 
    
    \item The experiments have been performed in both user-dependent as well as user-independent manner to ensure generalizability of the proposed airwriting recognition framework.
    
    \item All the source codes and the collected data used in this paper will be made available for usage in the HCI community. 
    
\end{itemize}

The remainder of the paper is organized as follows: Section II details the EMG data collection protocol (Section II.A), preprocessing steps (Section II.B), time-domain feature extraction (Section II.C), time-frequency feature extraction (Section II.D), and deep learning models(Section II.E). Experimental details and results are presented in Section III and Section IV concludes the paper.

\begin{figure*}[!t]
    \centering{
    \includegraphics[width=0.9\linewidth]{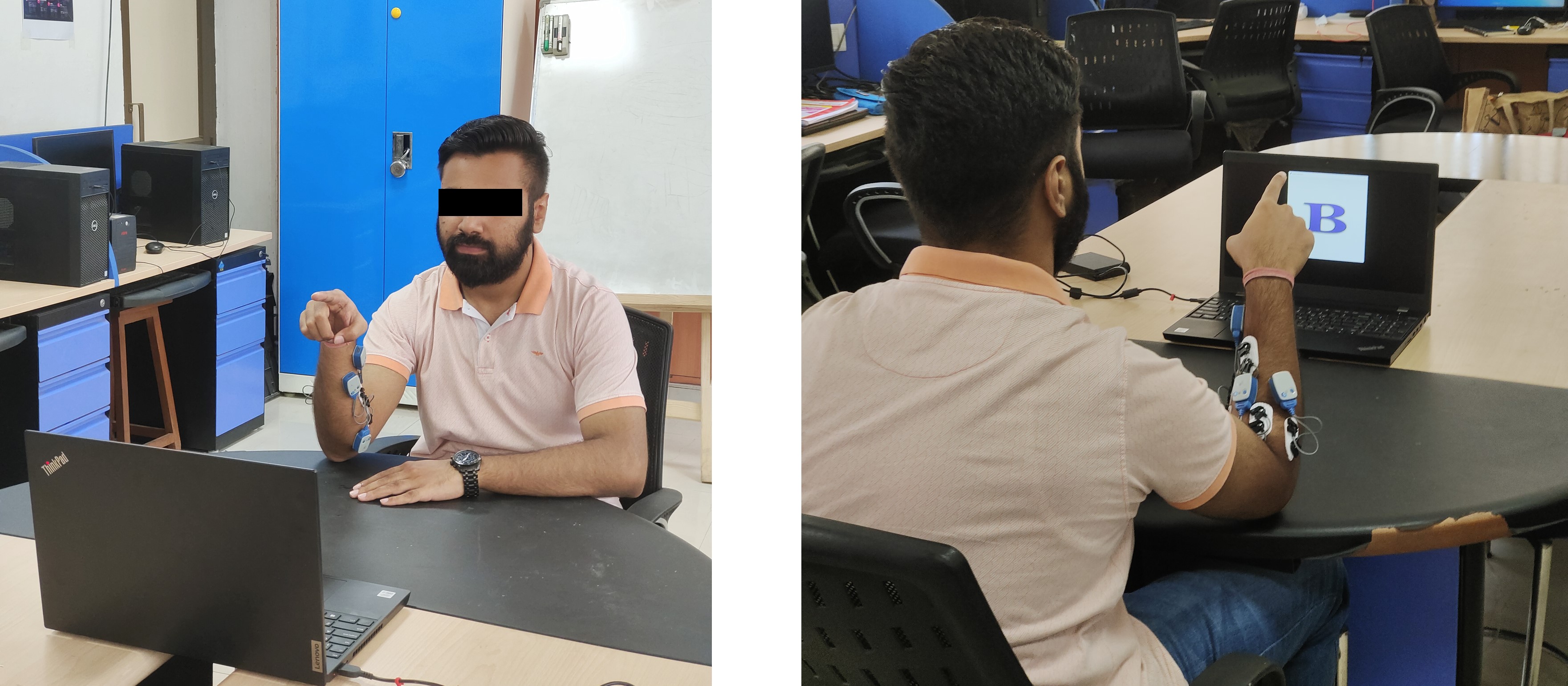}}
    \caption{Depiction of the data collection setup. The sEMG electrodes and sensors placed on forearm muscles of the subject and the Tkinter based user interface are depicted.}
    \label{fig:datacollection-a}
\end{figure*}

\begin{figure}[!t]
    \centering{
    \includegraphics[width=\linewidth]{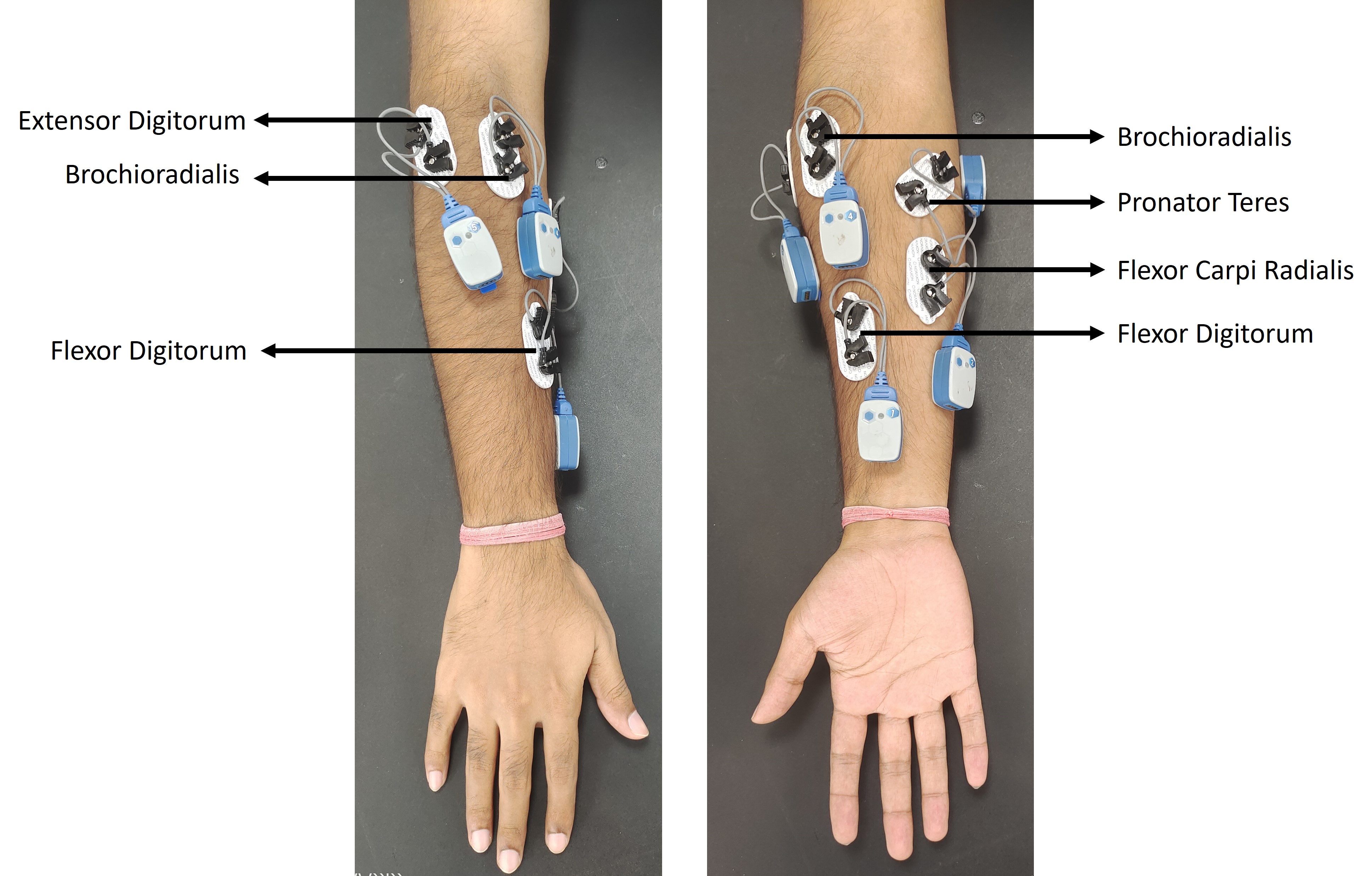}}
    \caption{Depiction of the sEMG electrode placement locations. The electrode orientation is in accordance with the muscle fibre orientation.}
    \label{fig:datacollection-b}
\end{figure}

\section{Material and Methods}

In this section, the sEMG data collection process, preprocessing of the raw sEMG signals, time-domain and time-frequency based feature extraction, and the Deep Learning models for classification are presented. 

\subsection{EMG Data Collection}

Surface EMG data for this work was collected in accordance with the guidelines of the Helsinki Declaration, and was approved by the Institute Ethics Committee, All India Institute of Medical Sciences, New Delhi, India. $50$ healthy subjects ($40$ male and $10$ female, all right-handed) with average age of $23.12$ years participated in the experiment. Before starting the recording session, the contents of the experiment were explained in detail to the participants and written consent was obtained. sEMG data was recorded from each participant's dominant hand. The signals were recorded using Noraxon Ultium wireless sEMG sensor \cite{noraxon} at sampling frequency of $2000$ Hz.  Disposable, wet-gel based, self-adhesive Ag/AgCl dual electrodes (having an inter-electrode spacing of $20 mm$) were placed on the skin over the target muscle according to the orientation of the muscle fibre. In order to keep the contact impedance low, the electrode placement location was cleaned with an alcohol solution. Based on the anatomical bony landmarks of the arm which is the standard method for surface EMG electrode placement, Pronator teres, Flexor Carpi Radialis, Flexor Digitorum, Extensor Digitorum, and Brachio Radialis were selected as the target muscles for our experiment. The electrode placement location was kept consistent for all the participants that took part in the study. 

The participant sat comfortably on a chair while firmly placing the elbow on the table. A user interface designed using the Tkinter module was used to give the participant visual cue regarding the character to be written and also to store annotations during the experiment. The user interface was operated by an experimenter overseeing the data recording session. On pressing the spacebar, a character randomly appeared on the screen and the participant was asked to write the character in a manner that they were writing on a whiteboard with their finger as the marker. In this manner, each of the participant wrote $10$ sets of $26$ English uppercase alphabets, thus resulting in $26\times10 = 260$ samples per subject and a total of $260\times50 = 13000$ samples. The alphabets within a set were randomly shuffled and did not appear in the usual alphabetical order. Furthermore, each alphabet was individually recorded and the subject was provided rest at the end of $2$ sets of recording. The duration of the rest duration was not fixed and the participant could take as much rest as required during this period. This was done to increase the variance between different repetitions of the same letter. The entire data collection setup and the sEMG electrode placement locations are presented in Figures \ref{fig:datacollection-a} and \ref{fig:datacollection-b} respectively.

\subsection{EMG Preprocessing}

\begin{figure}[!t]
    \centering{
    \includegraphics[width=\linewidth]{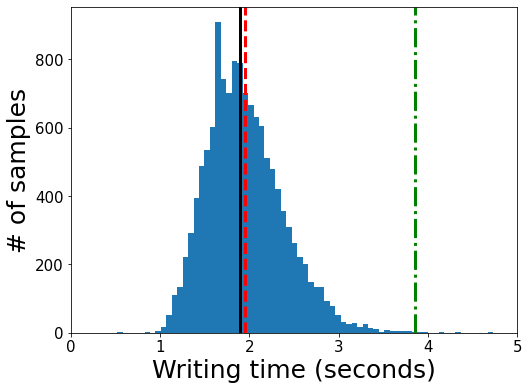}}
    \caption{Distribution of writing times of all samples across all subjects and alphabets. The red dashed line, black solid line and green dashed-dot lines depict the mean, median and $99.9$ percentile of the lengths respectively.}
    \label{fig:histogram_data}
\end{figure}

Given that the EMG signals are recorded from different users writing at different speeds and sizes, there exists large variation in the length of signals. Moreover, since different alphabets require different motion to be written, this adds to the variance in the signal lengths. A histogram depicting the duration of writing across all subjects and alphabets is presented in Figure \ref{fig:histogram_data}. The mean, median and $99.9$ percentile of duration taken to write an alphabet are $1.96$, $1.9$ and $3.86$ seconds respectively.  However, feeding the data to any deep learning architecture requires the input dimensions to be fix. In order to achieve this goal, interpolation is employed if the signal is of a length less than $L$, while discarding the extra samples otherwise. The effect of variation in the signal length, different values of the parameter $L \in \{2,2.5,3,3.5,4\}$ seconds is utilized in the experiments. Several standard one-dimensional interpolation techniques that include: linear, quadratic, cubic and nearest neighbor interpolation are utilized for this task. 
Suppose the length of a recorded sEMG signal is $l$ ($<L$), then, the first step is to define a list of $L$ time-stamps that are linearly spaced: $[0,\frac{1\cdot l}{L-1}, \frac{2\cdot l}{L-1},.....,l]$. Subsequently, for each time point, two values recorded just before and after this instant from the original signal are taken. Interpolation is then performed using one of the aforementioned techniques to obtain the value at the required time instant.  

\begin{figure*}[!ht]
\centering
\includegraphics[width=\textwidth]{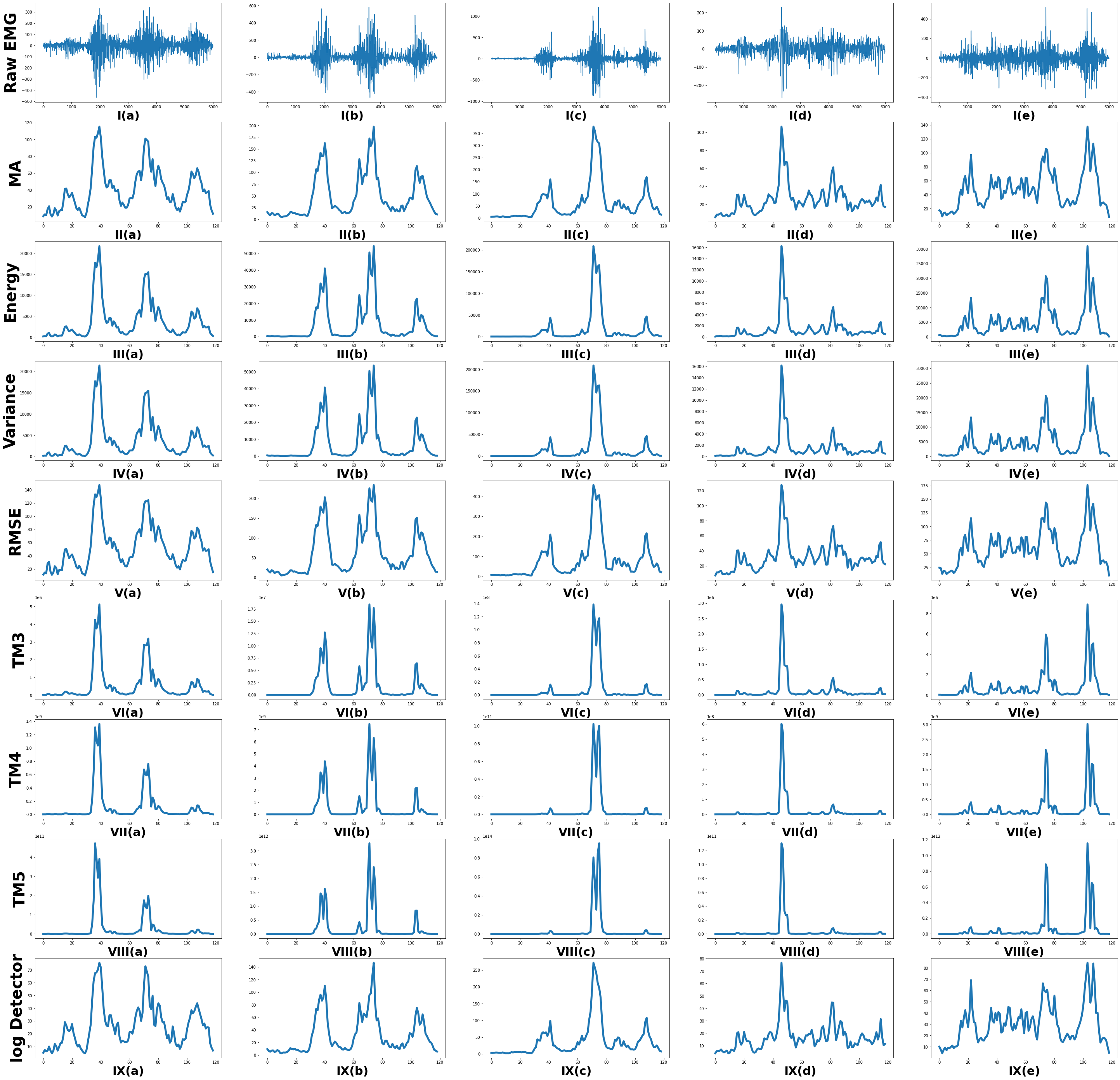}
\caption{Example of a five-channel EMG signal for the alphabet "A", interpolated to a length of 3 seconds ($6000$ samples) using linear interpolation (I(a)-I(e)), and the different envelopes extracted from the signal : Mean Absolute Envelope (II(a)-II(e)), Energy Envelope (III(a)-III(e)), Variance Envelope (IV(a)-IV(e)), Root Mean Squared Energy Envelope (V(a)-V(e)), Temporal Moment 3 Envelope (VI(a)-VI(e)), Temporal Moment 4 Envelope (VII(a)-VII(e)), Temporal Moment 5 Envelope (VIII(a)-VIII(e)), and log Detector Envelope (IX(a)-IX(e)).}
\label{fig:EMGEnvelopes}
\end{figure*}

\subsection{Time domain Analysis Approaches}

\begin{figure*}[!ht!]
\centering
\includegraphics[width=\linewidth]{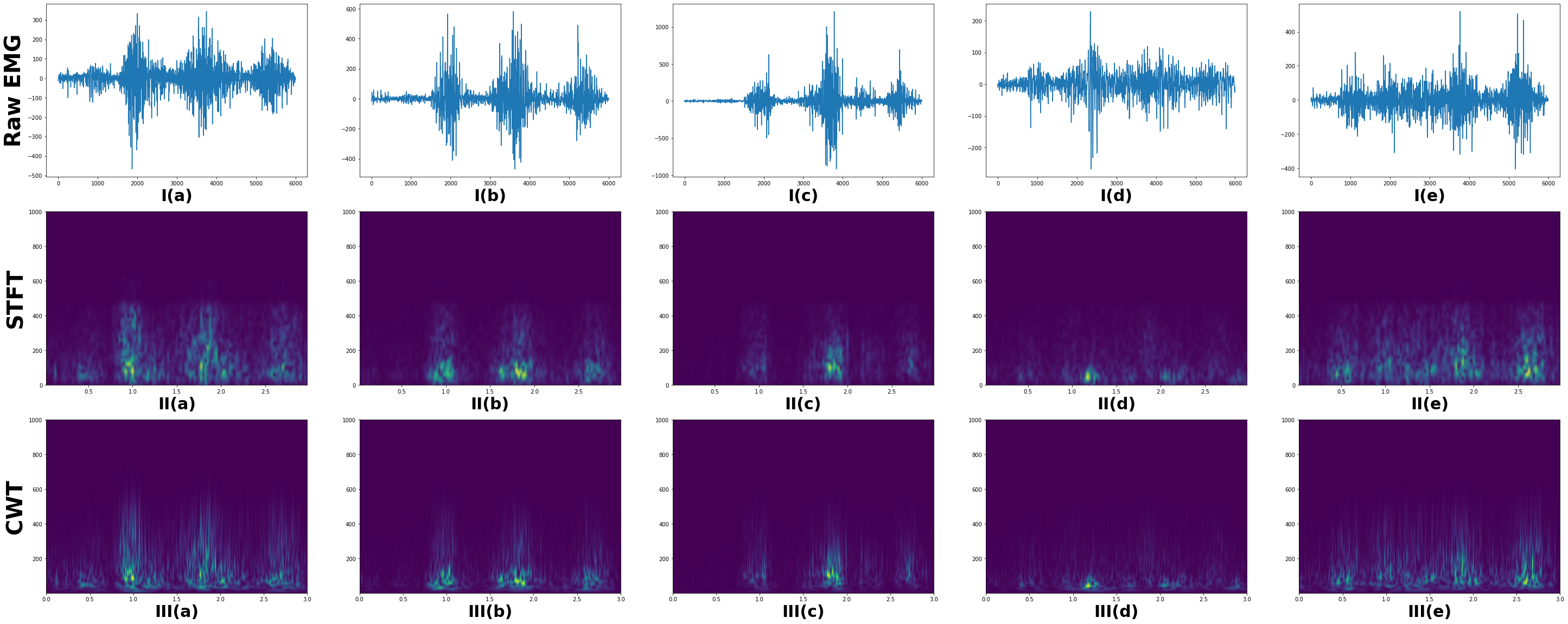}
\caption{Example of a five-channel EMG signal for the alphabet "A", interpolated to a length of 3 seconds ($6000$ samples) using linear interpolation (I(a)-I(e)), and the corresponding Time-Frequency images obtained by using Short-Time Fourier Transform (II(a)-II(e)) and Continuous Wavelet Transform (III(a)-III(e)).}
\label{fig:TF_domain}
\end{figure*}

To the fix length sEMG signals, sliding window method is applied to extract different sEMG envelopes from the raw signals. Features extracted from sliding rectangular windows of length W with an overlap of $50\%$ were used to construct envelopes from the $5$-channel EMG signals. The window length is varied with $W\in \{25,50,75,100,125,150\}$ millliseconds to comprehensively analyze the effect of variation of window length on the sEMG based airwriting recognition task. Several standard time-domain features were used for constructing the envelope which are depicted in Figure \ref{fig:EMGEnvelopes}. Given a windowed segment of signal from a single sEMG electrode $x$ of length $W$, the following methods have been applied to obtain the corresponding sEMG envelopes: 

\subsubsection{Mean Absolute Envelope}

The Mean Absolute Value (MAV) is an average of rectified sEMG amplitude within a segment and is used widely for EMG onset detection tasks. Mathematically, it is expressed as:

\begin{equation}
    MAV = \frac{1}{W} \sum_{i=1}^{W} |x_i|
\end{equation}

MAV provides an indication of the level of muscle contraction with the value being proportional to the amount of contraction. Additionally, it has also been used as a tool for detecting muscle movement onset.

\subsubsection{Energy Envelope}

Energy envelope is constructed by using the sum of square of the segmented sEMG signals as:

\begin{equation}
    Energy = \frac{1}{W} \sum_{i=1}^{W} x_i^2
\end{equation}

\subsubsection{Variance Envelope}

The variance of a signal segment is defined as the averaged square of deviation from the mean of the segment (denoted by $\mu_x$). It is a power index of the sEMG signal and computed as:

\begin{equation}
    Variance = \frac{1}{(W-1)} \sum_{i=1}^{W-1} (x_i - \mu_x)^2
\end{equation}

Since sEMG can be assumed as a zero mean process, the value $\mu_x$ can be taken to be $0$ for computing the variance feature.  

\subsubsection{Root Mean Square Envelope}

The Root Mean Square value is a measure of the strength of the segment and is mathematically computed as:

\begin{equation}
    RMS = \sqrt{\frac{1}{W} \sum_{i=1}^{W} x_i^2}
\end{equation}

It is an optimal method for estimating the standard deviation of a signal segment under the assumption of normal distribution. 

\subsubsection{Absolute Temporal Moment Envelope}
The first and second temporal moments are same same as Mean Absolute Value and Energy respectively while the $3$ different higher order temporal moments are computed as:

\begin{equation}
    TM3 = \frac{1}{W} \sum_{i=1}^{W} |x_i|^3
\end{equation}

\begin{equation}
    TM4 = \frac{1}{W} \sum_{i=1}^{W} x_i^4
\end{equation}

\begin{equation}
    TM5 = \frac{1}{W} \sum_{i=1}^{W} |x_i|^5
\end{equation}

Temporal moment of an sEMG signal has been widely used for applications in gesture recognition and prosthetic arm control in prior literature \cite{sharma2018use,saridis1982emg}. The absolute value taken during computation of a temporal moment is to ensure that the within class separation of the different movements is reduced.

\subsubsection{log Detector Envelope}

It is a non linear detector based on exponential-logarithm computation and is mathematically expressed as: 

\begin{equation}
    logD = exp\left({\frac{1}{W} \sum_{i=1}^{W} log(|x_i|)}\right)
\end{equation}

The log Detector feature provides an implicit estimation of the exerted muscle force and has been used in movement control of upper prostheses\cite{tkach2010study}.

\subsection{Time-Frequency Analysis Approaches}

Time Frequency Analysis is used to analyze the energy distribution of a signal jointly over both time and frequency domains. It is an effective way to convert the one-dimensional time series to a corresponding two-dimensional image representation. Several previous studies have utilized TF images for sEMG based classification tasks\cite{ozdemir2022hand,9119911}. In particular, two different approaches: Short-Time Fourier Transform and Continuous Wavelet Transform are investigated for the task of EMG based airwriting recognition (depicted in Figure \ref{fig:TF_domain}), which are briefly introduced in the following subsections. 

\subsubsection{Short-Time Fourier Transform}

STFT is a conventional approach used for representation and analysis of non-stationary signals. A sliding window is used to segment the signal and subsequently Fourier transform is computed for each segment which results in a joint time-frequency representation of the signal. Mathematically, for a signal $y(t)$ the STFT is computed as:

\begin{equation}
    Y[t,f] = \int_{-\infty}^{\infty} y(\tau) \omega(\tau - t) e^{-j2\pi f\tau}  \,d\tau\
\end{equation}

In this equation, $\omega(t)$ is a window function which is taken to be a hanning window. In practice, the Fast-Fourier Transform (FFT) is computed from a windowed sEMG signal while maintaining an overlap of $50\%$ between successive windows. The number of points for computing the FFT is taken to be same as the window length and one-sided spectrum is used for subsequent analysis. The STFT magnitude is further computed by taking the absolute value of the complex 2-dimensional signal $Y[t,f]$ which is fed to the deep learning model for the task of airwriting recognition.

\subsubsection{Continuous Wavelet Transform}

CWT is multi-resolution analysis technique that provides a representation of a signal by varying the translation and scale parameter of mother wavelets \cite{mallat1999wavelet}. The basis functions for CWT (denoted by $\Psi_{\sigma,\tau}(t)$)are described by scaling and translating a single mother wavelet function $\Psi(t)$. Mathematically, these basis functions can be represented as:

\begin{equation}
    \Psi_{\sigma,\tau}(t) = \frac{1}{\sqrt\sigma}\Psi\left(\frac{t-\tau}{\sigma}\right)
\end{equation}

In the aforementioned equation, $\tau$ is the translation parameter that shifts the mother wavelet over time and $\sigma$ is the scale factor. The normalizing factor $\frac{1}{\sqrt\sigma}$ ensures that the basis function has unit energy. In order to compute the CWT for a signal $y(t)$, the inner product of the signal with basis functions at different $\tau$ and $\sigma$ parameters is computed. Mathematically, it may be represented as:

\begin{equation}
    W_y^\Psi[\sigma,\tau] = y(t)\cdot\Psi_{\sigma,\tau}(t) = \frac{1}{\sqrt\sigma}\int_{-\infty}^{\infty} y(t) \Psi^*\left(\frac{t-\tau}{\sigma}\right)   \,dt\ 
\end{equation}

The wavelet coefficients are obtained by considering all possible shifts and scales of the Morlet mother wavelet. Subsequently the absolute value of the obtained CWT is taken, which is used for feeding to the deep learning model for classification.

\subsection{Deep Learning Frameworks}
Inspired by the success of deep learning models for solving classification tasks,several different architectures are used to curb and harness the task of airwriting recognition. The model architectures used for the sEMG envelope-based classification are detailed below. Extensive hyperparameter tuning was performed to identify the model architecture parameters. A detailed tabular description of all the models used in the study is provided in \cref{tab:Arch-1DCNN,tab:Arch-LSTM,tab:Arch-BiLSTM,tab:Arch-1DCNNLSTM,tab:Arch-1DCNNBiLSTM,tab:Arch-2DCNN}.

\begin{enumerate}
    \item 1DCNN: The 1D-CNN architecture is made up of $4$ convolutional layers, with a pooling layer sandwiched between pair of convolutional layers on either side. This is followed by a Global Average Pooling layer, a dropout layer and a $26$-neuron dense layer activated by the softmax activation function. In the first two convolution layers, the number of filters is set to be $100$ while it is set to $160$ in the last two convolution layers. The kernel size is set to $10$ for all the convolution layers and $3$ for the pooling layer. 
    
    \item LSTM/BiLSTM:  In this architecture, $2$ LSTM/BiLSTM layers each with $512$ number of units are stacked. This is followed by dropout and dense layers for classification.   
    
    \item 1DCNN-LSTM/BiLSTM: Here, the extracted sEMG envelope is first passed through two 1D convolution layers each having $200$ filters with a size of $10$. This is followed by a pooling layer with pool size of $3$. The features extracted from the CNN are subsequently fed to stacked LSTM/BiLSTM layers having $512$ units each. Subsequently, dropout and $26$-neuron dense layer activated by the softmax activation function are added. The effect of removing the pooling layer on the classification performance has also been explored. The models with and without the maxpooling layer are denoted by 1DCNN-Pool-LSTM/BiLSTM and 1DCNN-LSTM/BiLSTM respectively.
\end{enumerate}

In case of the models with LSTM/BiLSTM layers, the effect of adding the Attention mechanism \cite{bahdanau2014neural,vaswani2017attention} has also been explored. The attention mechanism aims at relating the different positions of a sequence to compute a weighted representation of the given sequence. 

For the time-frequency image based airwriting recognition, $2$ different schemes are adopted. In the first scenario, the time-frequency images for each of the $5$ channels are separated and fed to parallel models (with shared weights) having same configuration as that in the case of time-domain based classification. The feature vectors obtained from the $5$ parallel models are then concatenated, which is followed by dropout and dense layers. Additionally, a 2DCNN based model is also used for recognizing airwriting gestures from the time-frequency images. The 2DCNN architecture comprises of $4$ convolutional layers with the number of filters on each layer are the number of filters on the previous layer, starting with $32$ filters. Each convolutional layer is followed by a pooling layer and the output of the last layer is flattened and succeeded with a dropout and a dense layer. The kernel size of all convolution and pooling layers is set to $(3,3)$. 

\begin{table}[t!]
\caption{Details of the 1DCNN based classifier.}
\label{tab:Arch-1DCNN}
\centering
\scalebox{0.65}{
\begin{tabular}{cccc}
\hline
\textbf{Layer}  & \textbf{Kernel Size} & \textbf{\# of filters} & \textbf{Layer Parameters}                   \\\hline\hline
Conv1D          & 10                   & 100                    & Stride = 1, Activation = ReLU, Zero padding \\
Conv1D          & 10                   & 100                    & Stride = 1, Activation = ReLU, Zero padding \\
MaxPool1D       & 3                    &                        & Strides=3, No Padding                       \\
Conv1D          & 10                   & 160                    & Stride = 1, Activation = ReLU, Zero padding \\
Conv1D          & 10                   & 160                    & Stride = 1, Activation = ReLU, Zero padding \\
GlobalAvgPool1D & -                    & -                      & -                                           \\
Dropout         & -                    & -                      & Rate = 0.5                                  \\
Dense           & -                    & -                      & Neurons = 26, Activation = Softmax   \\\hline
\end{tabular}
}
\end{table}

\begin{table}[ht!]
\caption{Details of the LSTM based classifier.}
\label{tab:Arch-LSTM}
\centering
\scalebox{0.65}{
\begin{tabular}{cccc}
\hline
\textbf{Layer}  & \textbf{Kernel Size} & \textbf{\# of filters} & \textbf{Layer Parameters}                   \\\hline\hline
LSTM           & -                    & -                      & Hidden states = 512, Activation = tanh \\
LSTM           & -                    & -                      & Hidden states = 512, Activation = tanh \\
Dropout        & -                    & -                      & Rate = 0.5                             \\
Dense          & -                    & -                      & Neurons = 26, Activation = Softmax   \\\hline
\end{tabular}
}
\end{table}

\begin{table}[ht!]
\caption{Details of the Bidirectional LSTM based classifier.}
\label{tab:Arch-BiLSTM}
\centering
\scalebox{0.65}{
\begin{tabular}{cccc}
\hline
\textbf{Layer}  & \textbf{Kernel Size} & \textbf{\# of filters} & \textbf{Layer Parameters}                   \\\hline\hline
Bidirectional LSTM           & -                    & -                      & Hidden states = 512, Activation = tanh \\
Bidirectional LSTM           & -                    & -                      & Hidden states = 512, Activation = tanh \\
Dropout        & -                    & -                      & Rate = 0.5                             \\
Dense          & -                    & -                      & Neurons = 26, Activation = Softmax   \\\hline
\end{tabular}
}
\end{table}

\begin{table}[ht!]
\caption{Details of the 1DCNN-LSTM based classifier.}
\label{tab:Arch-1DCNNLSTM}
\centering
\scalebox{0.65}{
\begin{tabular}{cccc}
\hline
\textbf{Layer}  & \textbf{Kernel Size} & \textbf{\# of filters} & \textbf{Layer Parameters}                   \\\hline\hline
Conv1D         & 10                   & 100                    & Stride = 1, Activation = ReLU, Zero padding \\
Conv1D         & 10                   & 100                    & Stride = 1, Activation = ReLU, Zero padding \\
MaxPool1D      & 3                    &                        & Strides=3, No Padding                       \\
LSTM           & -                    & -                      & Hidden states = 512, Activation = tanh      \\
LSTM           & -                    & -                      & Hidden states = 512, Activation = tanh      \\
Dropout        & -                    & -                      & Rate = 0.5                                  \\
Dense          & -                    & -                      & Neurons = 26, Activation = Softmax   \\\hline
\end{tabular}
}
\end{table}

\begin{table}[ht!]
\caption{Details of the 1DCNN-BiLSTM based classifier.}
\label{tab:Arch-1DCNNBiLSTM}
\centering
\scalebox{0.65}{
\begin{tabular}{cccc}
\hline
\textbf{Layer}  & \textbf{Kernel Size} & \textbf{\# of filters} & \textbf{Layer Parameters}                   \\\hline\hline
Conv1D         & 10                   & 100                    & Stride = 1, Activation = ReLU, Zero padding \\
Conv1D         & 10                   & 100                    & Stride = 1, Activation = ReLU, Zero padding \\
MaxPool1D      & 3                    &                        & Strides=3, No Padding                       \\
BiLSTM           & -                    & -                      & Hidden states = 512, Activation = tanh      \\
BiLSTM           & -                    & -                      & Hidden states = 512, Activation = tanh      \\
Dropout        & -                    & -                      & Rate = 0.5                                  \\
Dense          & -                    & -                      & Neurons = 26, Activation = Softmax   \\\hline
\end{tabular}
}
\end{table}

\begin{table}[ht!]
\caption{Details of the 2DCNN based classifier.}
\label{tab:Arch-2DCNN}
\centering
\scalebox{0.65}{
\begin{tabular}{cccc}
\hline
\textbf{Layer}  & \textbf{Kernel Size} & \textbf{\# of filters} & \textbf{Layer Parameters}                   \\\hline\hline
Conv2D         & (3,3)                & 32                     & Stride = (1, 1), Activation = ReLU, Zero padding \\
MaxPool2D      & (2,2)                & -                      & Stride = (2,2), Zero padding                     \\
Conv2D         & (3,3)                & 64                     & Stride = (1, 1), Activation = ReLU, Zero padding \\
MaxPool2D      & (2,2)                & -                      & Stride = (2,2), Zero padding                     \\
Conv2D         & (3,3)                & 128                    & Stride = (1, 1), Activation = ReLU, Zero padding \\
MaxPool2D      & (2,2)                & -                      & Stride = (2,2), Zero padding                     \\
Conv2D         & (3,3)                & 256                    & Stride = (1, 1), Activation = ReLU, Zero padding \\
MaxPool2D      & (2,2)                & -                      & Stride = (2,2), Zero padding                     \\
Flatten        & -                    & -                      & -                                                \\
Dropout        & -                    & -                      & Rate = 0.5                                       \\
Dense          & -                    & -                      & Neurons = 26, Activation = Softmax   \\\hline
\end{tabular}
}
\end{table}

\section{Experimental Setup and Results}

\begin{table*}[!t]
\caption{Recognition accuracies for the airwriting recognition task by using different time-domain sEMG envelopes and model architecture combination in user-dependent and user-independent settings. Values are indicated as (mean $\pm$ standard deviation) of accuracies across the $5$ folds.}
\centering
\scalebox{0.8}{
\begin{tabular}{cccccccccc}
\hline
                     &                                & \multicolumn{8}{c}{\textbf{Approach}}                                                                                     \\\hline
                     & \textbf{Model}                 & \textbf{MA}  & \textbf{Energy} & \textbf{RMS} & \textbf{Var} & \textbf{TM3} & \textbf{TM4} & \textbf{TM5} & \textbf{logD} \\\hline\hline
                     & \textbf{1DCNN}                 & \textbf{75.75 $\pm$ 1.66} & 71.22 $\pm$ 2.31    & 75.10 $\pm$ 2.47 & 71.35 $\pm$ 1.40 & 63.93 $\pm$ 2.65 & 57.00 $\pm$ 1.26 & 50.63 $\pm$ 1.69 & 75.53 $\pm$ 1.92  \\
                     & \textbf{LSTM}                  & 74.79 $\pm$ 1.85 & 69.01 $\pm$ 1.86    & 73.77 $\pm$ 1.50 & 69.93 $\pm$ 1.51 & 61.57 $\pm$ 1.05 & 54.18 $\pm$ 2.10 & 46.92 $\pm$ 1.85 & 73.02 $\pm$ 2.13  \\
                     & \textbf{LSTM-Att}              & 75.34 $\pm$ 2.98 & 70.48 $\pm$ 2.33    & 73.84 $\pm$ 2.46 & 70.02 $\pm$ 3.04 & 63.02 $\pm$ 1.55 & 54.84 $\pm$ 2.95 & 47.62 $\pm$ 0.95 & 73.61 $\pm$ 2.05  \\
        & \textbf{BiLSTM}                & 74.24 $\pm$ 2.85 & 68.12 $\pm$ 2.53    & 72.12 $\pm$ 1.50 & 69.19 $\pm$ 1.58 & 60.59 $\pm$ 0.86 & 53.44 $\pm$ 2.36 & 46.54 $\pm$ 1.57 & 71.46 $\pm$ 2.94  \\
   & \textbf{BiLSTM-Att}            & 75.59 $\pm$ 2.34 & 70.18 $\pm$ 2.22    & 74.57 $\pm$ 2.83 & 70.68 $\pm$ 2.04 & 61.60 $\pm$ 2.26 & 55.42 $\pm$ 2.71 & 48.04 $\pm$ 2.54 & 74.43 $\pm$ 3.03  \\
        \textbf{User}             & \textbf{1DCNN-Pool-LSTM}       & 74.39 $\pm$ 3.04 & 70.02 $\pm$ 2.89    & 73.08 $\pm$ 2.43 & 68.65 $\pm$ 2.22 & 63.17 $\pm$ 1.65 & 54.12 $\pm$ 0.90 & 47.63 $\pm$ 1.17 & 72.69 $\pm$ 2.56  \\
         \textbf{Dependent}            & \textbf{1DCNN-Pool-LSTM-Att}   & 73.88 $\pm$ 2.39 & 68.05 $\pm$ 2.42    & 72.34 $\pm$ 2.23 & 68.51 $\pm$ 2.37 & 61.18 $\pm$ 2.55 & 52.72 $\pm$ 2.00 & 46.05 $\pm$ 2.78 & 71.99 $\pm$ 3.63  \\
                     & \textbf{1DCNN-LSTM}            & 74.54 $\pm$ 2.61 & 68.53 $\pm$ 1.72    & 72.84 $\pm$ 2.39 & 69.32 $\pm$ 2.02 & 61.78 $\pm$ 2.00 & 54.97 $\pm$ 2.10 & 49.94 $\pm$ 2.15 & 72.98 $\pm$ 2.26  \\
                     & \textbf{1DCNN-LSTM-Att}        & 73.78 $\pm$ 2.79 & 68.03 $\pm$ 1.60    & 72.35 $\pm$ 2.68 & 68.74 $\pm$ 1.82 & 60.86 $\pm$ 1.15 & 53.22 $\pm$ 1.94 & 46.75 $\pm$ 1.94 & 72.08 $\pm$ 2.58  \\
                     & \textbf{1DCNN-Pool-BiLSTM}     & 74.30 $\pm$ 1.85 & 68.62 $\pm$ 1.69    & 72.35 $\pm$ 1.66 & 67.89 $\pm$ 2.20 & 61.14 $\pm$ 1.76 & 52.60 $\pm$ 3.02 & 46.69 $\pm$ 2.73 & 72.62 $\pm$ 0.79  \\
                     & \textbf{1DCNN-Pool-BiLSTM-Att} & 72.89 $\pm$ 2.59 & 67.71 $\pm$ 1.54    & 71.56 $\pm$ 1.88 & 67.82 $\pm$ 2.49 & 60.21 $\pm$ 2.58 & 51.73 $\pm$ 2.20 & 45.54 $\pm$ 1.43 & 72.13 $\pm$ 2.23  \\
                     & \textbf{1DCNN-BiLSTM}          & 73.29 $\pm$ 2.67 & 68.70 $\pm$ 2.08    & 73.18 $\pm$ 3.32 & 68.75 $\pm$ 2.45 & 61.97 $\pm$ 1.49 & 54.06 $\pm$ 2.14 & 47.98 $\pm$ 1.02 & 72.28 $\pm$ 2.43  \\
                     & \textbf{1DCNN-BiLSTM-Att}      & 75.67 $\pm$ 3.21 & 68.98 $\pm$ 1.71    & 74.98 $\pm$ 3.02 & 69.90 $\pm$ 2.52 & 62.92 $\pm$ 2.29 & 54.37 $\pm$ 1.70 & 48.09 $\pm$ 1.76 & 73.18 $\pm$ 2.14  \\\hline\hline
                     & \textbf{1DCNN}                 & \textbf{61.13 $\pm$ 7.18} & 57.41 $\pm$ 7.00    & 59.77 $\pm$ 7.09 & 57.22 $\pm$ 6.94 & 52.12 $\pm$ 5.68 & 46.87 $\pm$ 6.13 & 41.69 $\pm$ 5.83 & 59.54 $\pm$ 8.31  \\
                     & \textbf{LSTM}                  & 57.45 $\pm$ 7.40 & 52.77 $\pm$ 7.84    & 56.56 $\pm$ 7.38 & 52.90 $\pm$ 7.64 & 46.81 $\pm$ 6.90 & 42.86 $\pm$ 5.75 & 37.12 $\pm$ 5.14 & 56.66 $\pm$ 7.82  \\
                     & \textbf{LSTM-Att}              & 57.78 $\pm$ 7.33 & 53.61 $\pm$ 7.25    & 56.20 $\pm$ 7.81 & 54.28 $\pm$ 6.34 & 47.58 $\pm$ 6.79 & 41.69 $\pm$ 5.85 & 37.16 $\pm$ 4.20 & 57.32 $\pm$ 7.97  \\
                     & \textbf{BiLSTM}                & 57.12 $\pm$ 7.14 & 52.60 $\pm$ 6.78    & 55.43 $\pm$ 6.53 & 52.52 $\pm$ 6.45 & 45.45 $\pm$ 6.37 & 40.96 $\pm$ 4.98 & 36.11 $\pm$ 4.41 & 56.58 $\pm$ 7.09  \\
                     & \textbf{BiLSTM-Att}            & 57.53 $\pm$ 6.68 & 53.52 $\pm$ 6.81    & 56.52 $\pm$ 8.33 & 53.86 $\pm$ 6.42 & 47.59 $\pm$ 6.15 & 42.82 $\pm$ 5.87 & 37.61 $\pm$ 5.42 & 56.05 $\pm$ 7.45  \\
\textbf{User}        & \textbf{1DCNN-Pool-LSTM}       & 58.64 $\pm$ 7.41 & 55.20 $\pm$ 7.02    & 57.39 $\pm$ 7.78 & 55.51 $\pm$ 7.26 & 48.42 $\pm$ 6.53 & 42.91 $\pm$ 5.67 & 38.22 $\pm$ 5.10 & 57.90 $\pm$ 7.77  \\
\textbf{Independent} & \textbf{1DCNN-Pool-LSTM-Att}   & 58.06 $\pm$ 8.14 & 54.09 $\pm$ 6.54    & 57.28 $\pm$ 7.54 & 53.69 $\pm$ 7.75 & 48.43 $\pm$ 6.51 & 41.45 $\pm$ 5.60 & 37.33 $\pm$ 5.43 & 57.13 $\pm$ 8.12  \\
                     & \textbf{1DCNN-LSTM}            & 57.49 $\pm$ 6.12 & 54.76 $\pm$ 6.60    & 56.35 $\pm$ 7.89 & 54.32 $\pm$ 6.12 & 48.70 $\pm$ 6.49 & 43.96 $\pm$ 5.50 & 39.99 $\pm$ 5.23 & 57.16 $\pm$ 7.19  \\
                     & \textbf{1DCNN-LSTM-Att}        & 58.98 $\pm$ 7.52 & 54.28 $\pm$ 8.05    & 58.26 $\pm$ 8.14 & 55.32 $\pm$ 7.27 & 49.62 $\pm$ 6.09 & 43.55 $\pm$ 4.68 & 38.74 $\pm$ 5.78 & 58.98 $\pm$ 6.70  \\
                     & \textbf{1DCNN-Pool-BiLSTM}     & 58.98 $\pm$ 7.30 & 54.39 $\pm$ 6.81    & 57.77 $\pm$ 7.53 & 53.16 $\pm$ 7.46 & 48.08 $\pm$ 6.87 & 42.45 $\pm$ 5.81 & 37.70 $\pm$ 5.05 & 56.85 $\pm$ 7.93  \\
                     & \textbf{1DCNN-Pool-BiLSTM-Att} & 57.59 $\pm$ 7.29 & 53.38 $\pm$ 6.95    & 57.20 $\pm$ 7.55 & 53.83 $\pm$ 6.73 & 47.73 $\pm$ 7.25 & 41.3 $\pm$ 6.14  & 36.47 $\pm$ 5.32 & 57.48 $\pm$ 7.58  \\
                     & \textbf{1DCNN-BiLSTM}          & 57.12 $\pm$ 6.96 & 53.65 $\pm$ 6.62    & 56.68 $\pm$ 7.66 & 53.45 $\pm$ 7.14 & 47.99 $\pm$ 6.43 & 43.16 $\pm$ 5.53 & 38.11 $\pm$ 5.39 & 56.68 $\pm$ 7.66  \\
                     & \textbf{1DCNN-BiLSTM-Att}      & 58.82 $\pm$ 7.91 & 55.12 $\pm$ 7.86    & 58.57 $\pm$ 7.60 & 55.11 $\pm$ 7.27 & 49.82 $\pm$ 6.38 & 42.82 $\pm$ 4.85 & 38.97 $\pm$ 4.82 & 57.39 $\pm$ 6.67 \\\hline
\end{tabular}
}
\label{tab:Results_timedomain}
\end{table*}

\begin{table}[!ht]
\caption{Recognition accuracies for the airwriting recognition task by using different time-frequency images and model architecture combination in user-dependent and user-independent settings. Values are indicated as (mean $\pm$ standard deviation) of accuracies across the $5$ folds.}
\centering
\scalebox{0.9}{
\begin{tabular}{cccc}
\hline
                     &                                & \multicolumn{2}{c}{\textbf{Approach}} \\\hline
                     & \textbf{Model}                 & \textbf{STFT}      & \textbf{CWT}     \\\hline\hline
                     & \textbf{1DCNN}                 & 75.11 $\pm$ 3.45       & 68.97 $\pm$ 2.70     \\
                     & \textbf{LSTM}                  & 69.90 $\pm$ 3.17       & 66.73 $\pm$ 3.05     \\
                     & \textbf{LSTM-Att}              & 73.25 $\pm$ 3.54       & 67.54 $\pm$ 3.05     \\
\multicolumn{1}{l}{} & \textbf{BiLSTM}                & 70.40 $\pm$ 2.47       & 65.82 $\pm$ 3.14     \\
\multicolumn{1}{l}{} & \textbf{BiLSTM-Att}            & 73.87 $\pm$ 2.24       & 68.76 $\pm$ 2.64     \\
\textbf{User}        & \textbf{1DCNN-Pool-LSTM}       & 71.65 $\pm$ 3.69       & 65.68 $\pm$ 2.56     \\
\textbf{Dependent}   & \textbf{1DCNN-Pool-LSTM-Att}   & 68.43 $\pm$ 4.77       & 63.68 $\pm$ 2.32     \\
                     & \textbf{1DCNN-LSTM}            & 71.67 $\pm$ 1.67       &  65.77 $\pm$ 1.87                \\
                     & \textbf{1DCNN-LSTM-Att}        & 71.01 $\pm$ 2.23       &      64.53 $\pm$ 3.12            \\
                     & \textbf{1DCNN-Pool-BiLSTM}     & 70.03 $\pm$ 2.13       & 66.03 $\pm$ 3.38     \\
                     & \textbf{1DCNN-Pool-BiLSTM-Att} & 71.44 $\pm$ 2.79       & 64.96 $\pm$ 4.18     \\
                     & \textbf{1DCNN-BiLSTM}          & 70.34 $\pm$ 3.50       &   64.92 $\pm$ 2.61               \\
                     & \textbf{1DCNN-BiLSTM-Att}      & 73.25 $\pm$ 3.26       &     67.31 $\pm$ 2.12             \\
                     & \textbf{2DCNN}                 & \textbf{78.50 $\pm$ 2.33}       & 72.98 $\pm$ 2.16     \\\hline\hline
                     & \textbf{1DCNN}                 & 55.84 $\pm$ 5.57       & 53.18 $\pm$ 6.15     \\
                     & \textbf{LSTM}                  & 52.38 $\pm$ 6.21       & 48.78 $\pm$ 6.11     \\
                     & \textbf{LSTM-Att}              & 54.82 $\pm$ 5.95       & 50.85 $\pm$ 6.43     \\
\multicolumn{1}{l}{} & \textbf{BiLSTM}                & 50.72 $\pm$ 5.86       & 48.85 $\pm$ 5.79     \\
\multicolumn{1}{l}{} & \textbf{BiLSTM-Att}            & 54.68 $\pm$ 7.41       & 49.88 $\pm$ 5.66     \\
\textbf{User}        & \textbf{1DCNN-Pool-LSTM}       & 54.25 $\pm$ 6.11       & 49.62 $\pm$ 6.59     \\
\textbf{Independent} & \textbf{1DCNN-Pool-LSTM-Att}   & 52.25 $\pm$ 5.57       & 49.34 $\pm$ 6.37     \\
                     & \textbf{1DCNN-LSTM}            & 52.99 $\pm$ 5.83       &   50.22 $\pm$ 6.38               \\
                     & \textbf{1DCNN-LSTM-Att}        & 52.20 $\pm$ 5.45       &    49.28 $\pm$ 5.41              \\
                     & \textbf{1DCNN-Pool-BiLSTM}     & 51.88 $\pm$ 6.73       & 49.75 $\pm$ 6.65     \\
                     & \textbf{1DCNN-Pool-BiLSTM-Att} & 51.78 $\pm$ 6.27       & 49.34 $\pm$ 5.59     \\
                     & \textbf{1DCNN-BiLSTM}          & 52.08 $\pm$ 6.13       &    48.55 $\pm$ 6.12              \\
                     & \textbf{1DCNN-BiLSTM-Att}      & 52.95 $\pm$ 5.24       &    50.96 $\pm$ 6.00              \\
                     & \textbf{2DCNN}                 & \textbf{62.19 $\pm$ 7.59}       & 58.15 $\pm$ 7.15    \\\hline
\end{tabular}
}
\label{tab:Results_tfdomain}
\end{table}

\subsection{Experimental Details}

The recorded sEMG signals from different users correspond to letters written at different speed and sizes. This leads to a large variation in amplitude of the sEMG signals, and subsequently the extracted envelopes and time-frequency representations. Therefore, z-normalization is applied to each channel of the obtained sEMG envelopes and time-frequency images. In order to check for the robustness of the proposed airwriting recognition system, two different validation schemes are adopted. 

\begin{figure*}[!t]
\centering
\includegraphics[width=0.8\linewidth]{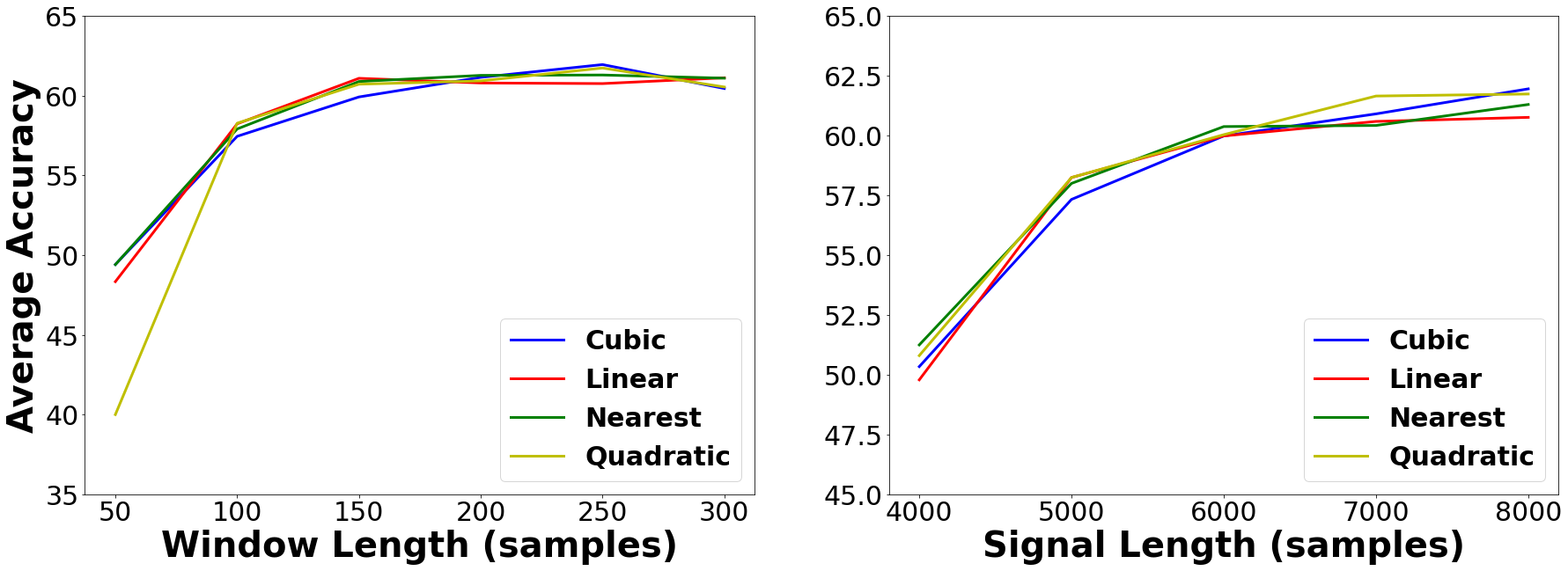}
\caption{Effect of variation of different parameters on mean recognition accuracies for the airwriting recognition system using MA Envelope and 1DCNN model. The left plot depicts the variation of accuracy with window length while keeping the signal length fix to $8000$ samples (4 seconds), while the plot on the right depicts the variation of accuracy with signal length while keeping the window length fix to $250$ samples (125 milliseconds). Effect of different interpolation techniques is marked by the legend in the subplot.}
\label{fig:parameters-a}
\end{figure*}

\begin{figure*}[!t]
\centering
\includegraphics[width=0.8\linewidth]{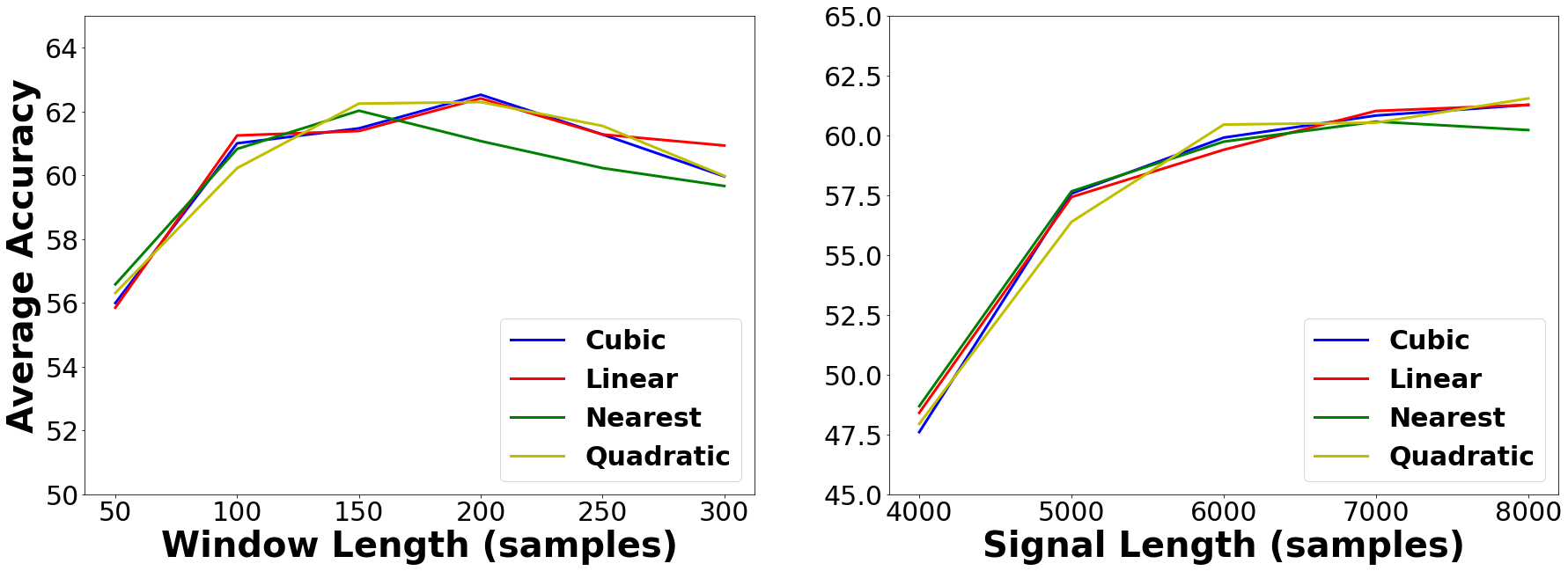}
\caption{Effect of variation of different parameters on mean recognition accuracies for the airwriting recognition system using STFT and 2DCNN model. The left plot depicts the variation of accuracy with window length while keeping the signal length fix to $8000$ samples (4 seconds), while the plot on the right depicts the variation of accuracy with signal length while keeping the window length fix to $250$ samples (125 milliseconds). Effect of different interpolation techniques is marked by the legend in the subplot.}
\label{fig:parameters-b}
\end{figure*}

\begin{enumerate}
    \item User-independent validation: In this setting, $5$-fold validation is performed while ensuring that there is no subject overlap in the training and test sets. For this purpose, data corresponding to $40$ subjects ($40*26*10 = 10400$ samples) is used for training the model and data from remaining $10$ subjects ($10*26*10 = 2600$ samples) is used for testing the performance of the trained model. This process is then repeated $5$ times to ensure that accuracy on each subject is evaluated. 
    
    \item User-dependent validation: In this setting, the train-test split is done by dividing the entire data based on the repetition number during the airwriting data collection process. Training set is made by combining data from $8$ (out of the $10$) repetitions from all subjects and the remaining $2$ repetitions form the test set. Similar to the user-independent setting, the process is repeated $5$ times to ensure that each repetition is a part of test set in one of the folds.
\end{enumerate}

The training data in each fold is further split into a training and a validation set in a $80:20$ ratio. The parameters of the model are learnt by minimizing the categorical cross-entropy loss. A mini-batch training process with a batch size of $256$ is employed and the parameters of the model are updated using the Adam optimizer\cite{kingma2014adam}. To prevent overfitting of the model, in addition to the $50\%$ dropout, early stopping with a patience of $10$ epochs while monitoring the accuracy on the validation set is applied.

\subsection{Results}

Table \ref{tab:Results_timedomain} lists the recognition accuracies by using different time-domain sEMG envelopes and deep learning models. For all the results presented in the table, the sEMG signals are interpolated to $4$ second duration ($8000$ samples) by using the cubic interpolation technique. Subsequently, sEMG envelope is constructed by using a window size of $125$ milliseconds ($250$ samples). In Table \ref{tab:Results_tfdomain}, the mean recognition accuracy for the airwriting recognition task by using the time-frequency images is presented. The STFT image is constructed with a window size of $100$ milliseconds ($200$ samples) after cubic interpolation of sEMG signals to $4$ seconds ($8000$ samples), while CWT is computed at $60$ scale values. 

\begin{figure*}
\centering  
\subfigure[]{\includegraphics[width=\linewidth]{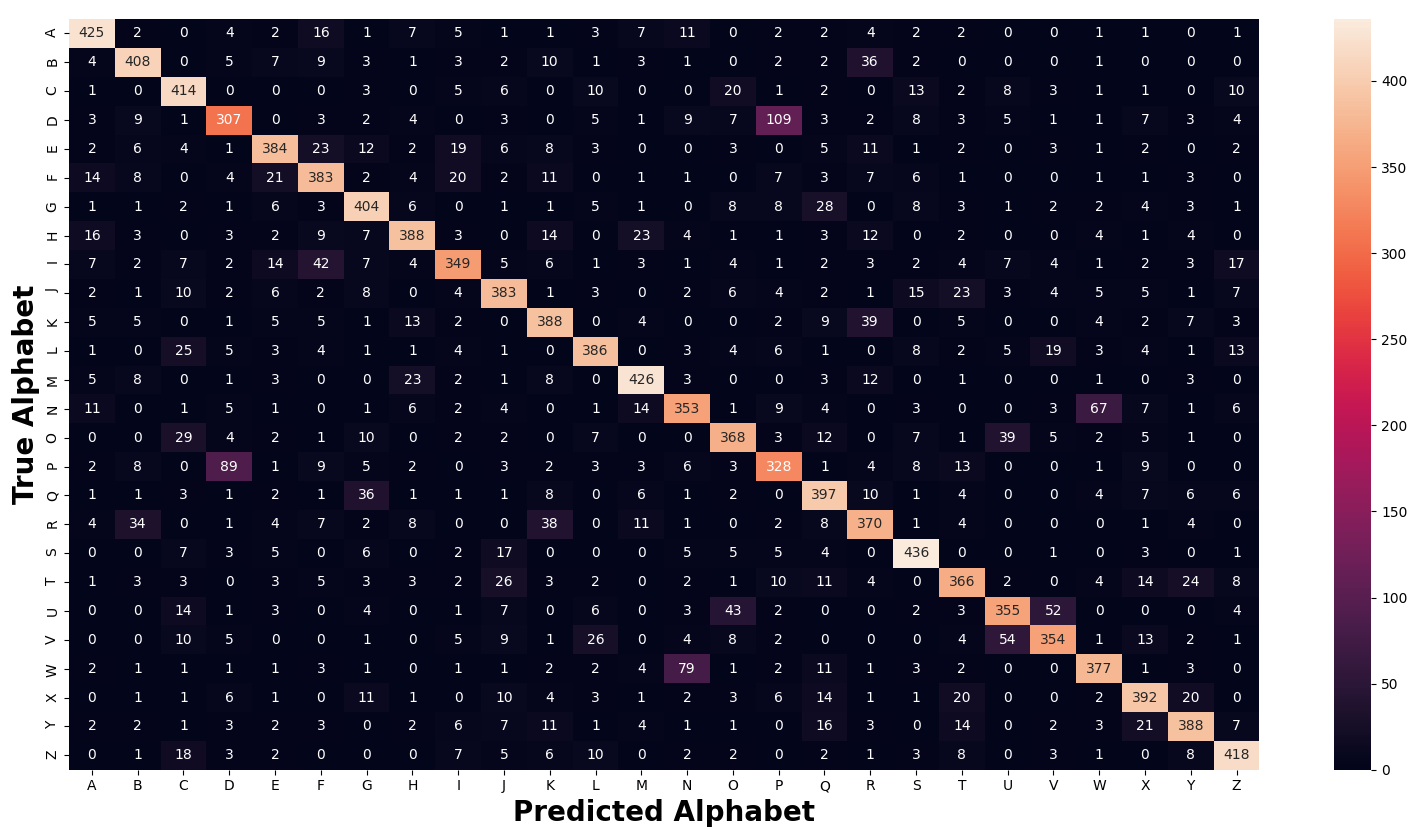}}
\subfigure[]{\includegraphics[width=\linewidth]{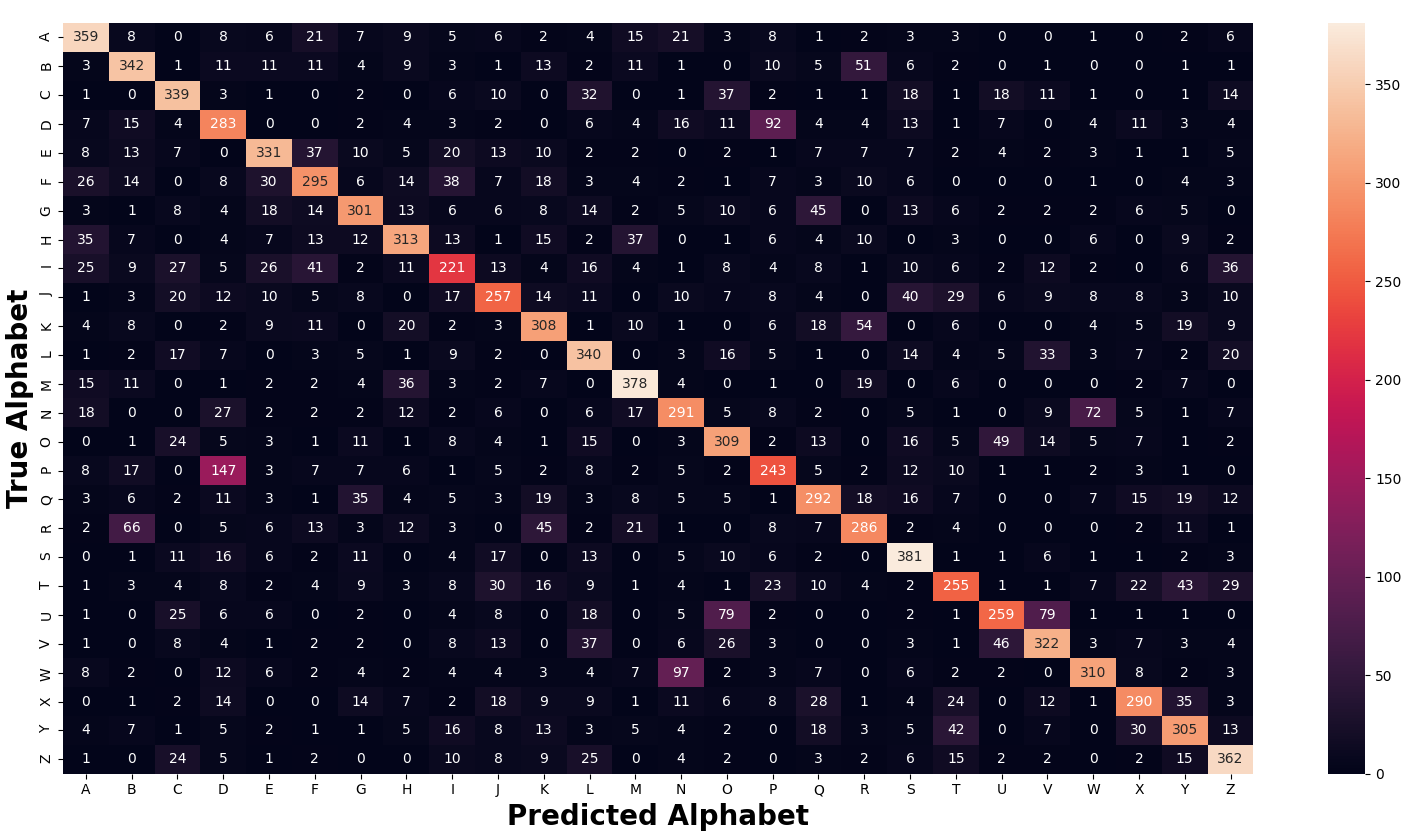}}
\caption{Confusion matrices corresponding to the MA Envelope and 1DCNN model in (a) user-dependent, and (b) user-independent scenarios.}
\label{fig:ConfMat}
\end{figure*}

The effect of varying different parameters such as window length, signal length and interpolation techniques on the airwriting recognition performance using Mean Absolute Envelope with the 1DCNN model based classification scheme is presented in Figure \ref{fig:parameters-a}. The corresponding variation using STFT images and 2DCNN model is presented in Figure \ref{fig:parameters-b}. Additionally, to analyze the classification results closely, the confusion matrices corresponding to MA envelope and 1DCNN model architecture are presented in Figure \ref{fig:ConfMat}. From the figure, the letter pairs that lead to the highest contribution to the total misclassification can be observed. The top-$5$ such pairs in for both user-independent and user-dependent approaches are presented in Tables \ref{tab:ConfusingLetters_ma_userindep} and \ref{tab:ConfusingLetters_ma_userdep} respectively.

\begin{table}[!ht]
\caption{The $5$ most confusing alphabet pairs for MA Envelope and 1DCNN model in the user-independent setting.}
\label{tab:ConfusingLetters_ma_userindep}
\centering
\scalebox{1.0}{
\begin{tabular}{ccc}
\hline
\textbf{Rank} & \textbf{Letter Pair} & \textbf{\% of Total Error} \\\hline\hline
1             & D,P                                       & 4.75\%                                        \\
2             & N,W                                       & 3.36\%                                         \\
3             & O,U                                       & 2.55\%                                         \\
4             & U,V                                       & 2.49\%                                         \\
5             & B,R                                       & 2.33\%   \\ \hline                                     
\end{tabular}
}
\end{table}%

\begin{table}[!ht]
\caption{The $5$ most confusing alphabet pairs for MA Envelope and 1DCNN model in the user-dependent setting.}
\label{tab:ConfusingLetters_ma_userdep}
\centering
\scalebox{1.0}{
\begin{tabular}{ccc}
\hline
\textbf{Rank} & \textbf{Letter Pair} & \textbf{\% of Total Error} \\\hline\hline
1             & D,P                                       & 6.48\%                                        \\
2             & N,W                                       & 4.78\%                                         \\
3             & U,V                                       & 3.47\%                                         \\
4             & O,U                                      & 2.68\%                                         \\
5             & K,R                                      & 2.52\%   \\ \hline                                     
\end{tabular}
}
\end{table}

\subsection{Discussion}

From Table \ref{tab:Results_timedomain}, it may be noted that using the mean absolute sEMG envelope coupled with 1DCNN based model architecture yields the best recognition accuracy of $75.75\%$ and $61.13\%$ in user-dependent and user-independent scenarios. The superior performance of the mean absolute envelope may be attributed to the its capability of capturing the fine details contained in the sEMG signal compared to other envelope construction strategies. This is also evident from the sample envelopes presented in Figure \ref{fig:EMGEnvelopes}. Likewise, the envelopes constructed by using root mean square and log Detector features yield a comparable performance to the mean absolute envelope due to the aforementioned reason.  It may also be noted that, on using higher order features such as energy and temporal moments, the fine-level details of the the sEMG signal are lost which leads to a significant reduction in the classification performance. Among the different classifiers, the 1DCNN model outperforms all the other architectures due to its ability to exploit the spatial correlation in the sEMG envelopes and hence suiting well for this task. The other deep learning architectures used in the study also perform fairly well, yielding comparable recognition accuracies. Performance of LSTM/BiLSTM and 1DCNN-LSTM/BiLSTM models is boosted by using the Attention mechanism. However, in case of the 1DCNN-Pool-LSTM/BiLSTM model, employing Attention does not improve the model performance since the signal dimension gets shortened due to the pooling layer. On using the time-frequency images, best accuracy is achieved on using the STFT and 2DCNN model architecture combination. An accuracy of $78.50\%$ and $62.19\%$ is achieved in user-independent and user-dependent settings respectively. The representation of the sEMG signal in joint time and frequency space helps the 2DCNN model to extract diverse characteristics and learn the complex patterns which is helpful for the airwriting recognition task. It is to be noted that airwriting is a dynamic gesture and the gesture vocabulary comprises of $26$ classes (chance accuracy of $3.84\%$). This implies that the accuracies achieved by the models are well within acceptable limits for practical usage. A single-factor Analysis of Variance (ANOVA) performed on the accuracies obtained from the $5$-folds by using different sEMG envelopes with the 1DCNN based classification model yielded p-values of $0.00047$ and $8.21\times10^{-21}$ in user-independent and dependent settings respectively. Similarly, p-values of $0.2054$ and $0.0082$ were obtained by using the 2DCNN model with the time-frequency based approach in user-independent and dependent settings by performing one tailed t-test. This implies that there is a statistically significant difference between the results obtained by using different features for the task of airwriting recognition using sEMG signals. A detailed t-test between all possible pairs of sEMG envelopes with 1DCNN model was also performed and the results are presented in Tables \ref{tab:ttest-userindep-timeseries} and \ref{tab:ttest-userdep-timeseries}.    

\begin{table}[h]
\caption{p-value of the comparative one tailed t-test between different time domain envelopes and the 1DCNN based classifier in user-independent setting. Entries with a significance value below 0.05 are depicted in blue.}
\label{tab:ttest-userindep-timeseries}
\centering
\scalebox{0.75}{
\begin{tabular}{|c|c|c|c|c|c|c|c|c|}
\hline
       & MA                            & Energy                        & RMS                           & Var                           & TM3                           & TM4                           & TM5                           & logD                          \\ \hline
MA     & -                             & 0.2151                        & 0.3852                        & 0.2029                        & {\color[HTML]{0000FF} 0.0294} & {\color[HTML]{0000FF} 0.0048} & {\color[HTML]{0000FF} 0.0008} & 0.3770                        \\ \hline
Energy & 0.2151                        & -                             & 0.3053                        & 0.4831                        & 0.1128                        & {\color[HTML]{0000FF} 0.0176} & {\color[HTML]{0000FF} 0.0024} & 0.3363                        \\ \hline
RMS    & 0.3852                        & 0.3053                        & -                             & 0.2904                        & {\color[HTML]{0000FF} 0.0482} & {\color[HTML]{0000FF} 0.0076} & {\color[HTML]{0000FF} 0.0011} & 0.4818                        \\ \hline
Var    & 0.2029                        & 0.4831                        & 0.2904                        & -                             & 0.1196                        & {\color[HTML]{0000FF} 0.0185} & {\color[HTML]{0000FF} 0.0025} & 0.3221                        \\ \hline
TM3    & {\color[HTML]{0000FF} 0.0294} & 0.1128                        & {\color[HTML]{0000FF} 0.0482} & 0.1196                        & -                             & 0.0991                        & {\color[HTML]{0000FF} 0.0105} & 0.0716                        \\ \hline
TM4    & {\color[HTML]{0000FF} 0.0048} & {\color[HTML]{0000FF} 0.0176} & {\color[HTML]{0000FF} 0.0076} & {\color[HTML]{0000FF} 0.0185} & 0.0991                        & -                             & 0.1042                        & {\color[HTML]{0000FF} 0.0144} \\ \hline
TM5    & {\color[HTML]{0000FF} 0.0008} & {\color[HTML]{0000FF} 0.0024} & {\color[HTML]{0000FF} 0.0011} & {\color[HTML]{0000FF} 0.0025} & {\color[HTML]{0000FF} 0.0105} & 0.1042                        & -                             & {\color[HTML]{0000FF} 0.0028} \\ \hline
logD   & 0.3770                        & 0.3363                        & 0.4818                        & 0.3221                        & 0.0716                        & {\color[HTML]{0000FF} 0.0144} & {\color[HTML]{0000FF} 0.0028} & -                             \\ \hline
\end{tabular}
}
\end{table}

\begin{table}[h]
\caption{p-value of the comparative one tailed t-test between different time domain envelopes and the 1DCNN based classifier in user-dependent setting. Entries with a significance value below 0.05 are depicted in blue.}
\label{tab:ttest-userdep-timeseries}
\centering
\scalebox{0.5}{
\begin{tabular}{|c|c|c|c|c|c|c|c|c|}
\hline
       & MA                                 & Energy                             & RMS                                & Var                                & TM3                                & TM4                                & TM5                                & logD                               \\ \hline
MA     & -                                  & {\color[HTML]{0000FF} 0.0046}      & 0.3193                             & {\color[HTML]{0000FF} 0.0009}      & {\color[HTML]{0000FF} 3.18$\times10^{-5}$} & {\color[HTML]{0000FF} 9.41$\times10^{-8}$} & {\color[HTML]{0000FF} 5.31$\times10^{-9}$} & 0.4246                             \\ \hline
Energy & {\color[HTML]{0000FF} 0.0046}      & -                                  & {\color[HTML]{0000FF} 0.0168}      & 0.4608                             & {\color[HTML]{0000FF} 0.0008}      & {\color[HTML]{0000FF} 9.68$\times10^{-6}$}  & {\color[HTML]{0000FF} 4.32$\times10^{-7}$} & {\color[HTML]{0000FF} 0.0062}      \\ \hline
RMS    & 0.3193                             & {\color[HTML]{0000FF} 0.0168}      & -                                  & {\color[HTML]{0000FF} 0.0128}      & {\color[HTML]{0000FF} 6.25$\times10^{-5}$} & {\color[HTML]{0000FF} 3.26$\times10^{-6}$} & {\color[HTML]{0000FF} 1.81$\times10^{-7}$} & 0.3832                             \\ \hline
Var    & {\color[HTML]{0000FF} 0.0009}      & 0.4608                             & {\color[HTML]{0000FF} 0.0128}      & -                                  & {\color[HTML]{0000FF} 0.0007}      & {\color[HTML]{0000FF} 7.16$\times10^{-8}$} & {\color[HTML]{0000FF} 1.32$\times10^{-8}$}  & {\color[HTML]{0000FF} 0.0028}      \\ \hline
TM3    & {\color[HTML]{0000FF} 3.18$\times10^{-5}$} & {\color[HTML]{0000FF} 0.0008}      & {\color[HTML]{0000FF} 6.25$\times10^
{-5}$} & {\color[HTML]{0000FF} 0.0007}      & -                                  & {\color[HTML]{0000FF} 0.0009}      & {\color[HTML]{0000FF} 1.52$\times10^{-5}$} & {\color[HTML]{0000FF} 4.82$\times10^{-5}$} \\ \hline
TM4    & {\color[HTML]{0000FF} 9.41$\times10^{-8}$} & {\color[HTML]{0000FF} 9.68$\times10^{-6}$}  & {\color[HTML]{0000FF} 3.26$\times10^{-6}$} & {\color[HTML]{0000FF} 7.16$\times10^{-8}$} & {\color[HTML]{0000FF} 0.0009}      & -                                  & {\color[HTML]{0000FF} 0.0001}      & {\color[HTML]{0000FF} 1.99$\times10^{-7}$}  \\ \hline
TM5    & {\color[HTML]{0000FF} 5.31$\times10^{-9}$} & {\color[HTML]{0000FF} 4.32$\times10^{-7}$} & {\color[HTML]{0000FF} 1.81$\times10^{-7}$} & {\color[HTML]{0000FF} 1.32$\times10^{-8}$}  & {\color[HTML]{0000FF} 1.52$\times10^{-5}$} & {\color[HTML]{0000FF} 0.0001}      & -                                  & {\color[HTML]{0000FF} 1.04$\times10^{-8}$} \\ \hline
logD   & 0.4246                             & {\color[HTML]{0000FF} 0.0062}      & 0.3832                             & {\color[HTML]{0000FF} 0.0028}      & {\color[HTML]{0000FF} 4.82$\times10^{-5}$} & {\color[HTML]{0000FF} 1.99$\times10^{-7}$}  &  {\color[HTML]{0000FF} 1.04$\times10^{-8}$}                                  & -                                  \\ \hline
\end{tabular}
}
\end{table}

From Figures \ref{fig:parameters-a} and \ref{fig:parameters-b}, it may be observed that on increasing the length of the signal, the performance of the airwriting recognition system improves. This increasing trend is due to the fact that discarding samples to constraint the signal length leads to a loss of information which is vital for the model to learn nuances of the airwritten letter. The choice of window length parameter is influenced by a trade-off between smoothness of the sEMG signal and capturing the fine details of the sEMG signal. Hence, the accuracy at first increases with increasing window size but then converges (and drops slightly) upon further increment. Therefore, a window size of 125 milliseconds (250 samples) was chosen as optimum for all the analysis. Similar trade-off between time and frequency resolution of the resulting STFT image yields a window size of 100 milliseconds (200 samples) as optimum for obtaining the time-frequency image.   

A list of top-5 pairs with highest contribution to the error in user-independent and user-dependent settings by using the mean absolute envelope and 1DCNN model is presented in Table \ref{tab:ConfusingLetters_ma_userindep} and \ref{tab:ConfusingLetters_ma_userdep} respectively. It may be noted that letters pairs with visible similarity contribute the highest to the misclassification. For instance, 'D' and 'P' are the most widely confusing letter pairs which is attributed to the fact that both the alphabets involve similar movement required to write them in the air (vertical line and semi-circle). A user can easily lose spatial orientation while writing the letters thereby leading to high misclassification rate.

\section{Conclusions}

In this paper, an sEMG based airwriting recognition framework was proposed. The SurfMyoAiR dataset comprising of sEMG signals measured from $5$ forearm muscles of $50$ subjects while writing the $26$ English uppercase alphabets was constructed. To the best of our knowledge, this is the first large-scale dataset for the airwriting recognition task. Several sEMG envelope extraction methods using time-domain features such as: mean absolute value, energy, variance, root mean square, different temporal moments and log Detector, and time-frequency images: Short-Time Fourier Transform and Continuous Wavelet Transform were explored to form input to deep learning models for airwriting recognition. The effect of different parameters such as signal length, window length and different interpolation techniques on the system performance in both user-dependent and user-independent settings was also comprehensively examined. Among the different EMG envelopes, the mean absolute envelope yielded the best recognition accuracy of $61.13\%$ and $75.75\%$ in the user independent and dependent settings respectively. This is attributed to the fact that the mean absolute envelope captures finer details of the EMG signal which leads a superior representation compared to other envelope construction approaches. The STFT image outperforms all the techniques explored in the paper giving a recognition accuracy of $62.19\%$ and $78.50\%$ in the user independent and dependent settings respectively. This is expected as complex attributes corresponding to different airwritten letters are better observable in a joint time-frequency space. Future
work may be focused on improving the performance of the system by exploring other sophisticated feature extraction and/or deep learning techniques. In conclusion, the fairly high recognition accuracies form a great baseline for future work in the domain of sEMG based airwriting recognition. Such an approach has a potential to be used as an alternate input method for Human Computer Interaction applications.

\section*{Acknowledgment}

The authors would like to thank Shivani Ranjan, Anant Jain and Mirdoddi Kaushik for their valuable insights and assistance during data collection and all the subjects for their participation in the experiment.

\bibliographystyle{IEEEtran}
\bibliography{refs.bib}

\begin{IEEEbiography}
[{\includegraphics[width=1in,height=1.25in,clip,keepaspectratio]{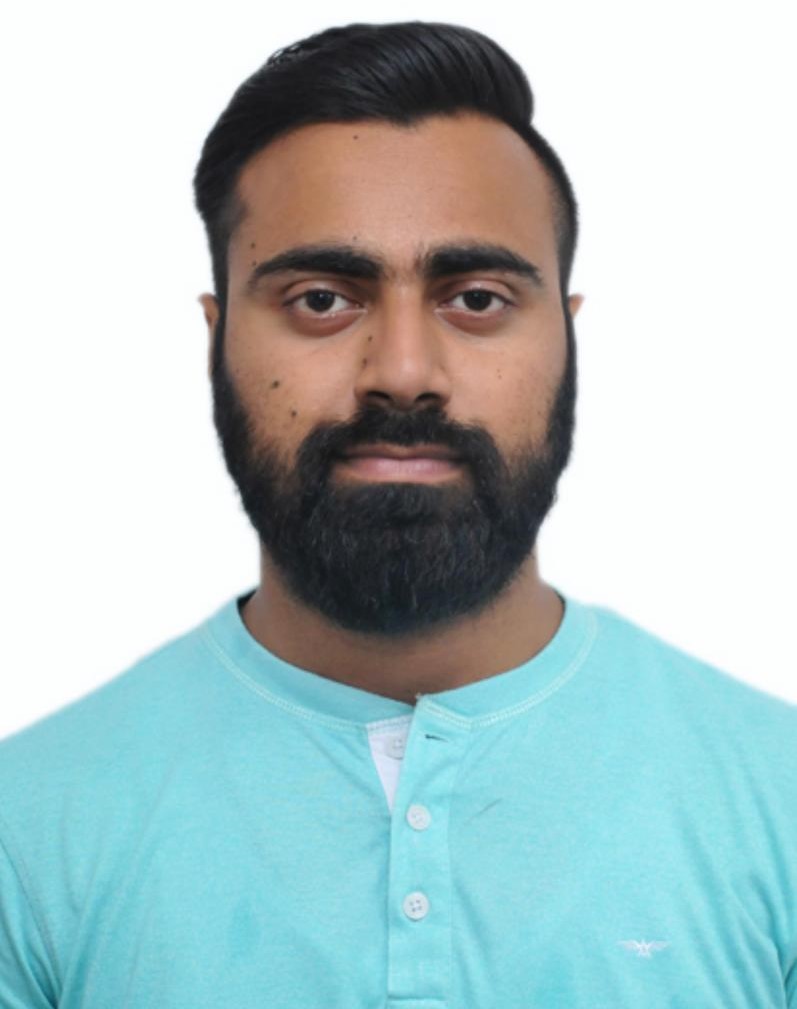}}]
{Ayush Tripathi} received the B.Tech in Electrical Engineering from Visvesvaraya National Institute of Technology, Nagpur, India, in 2019. During his B.Tech, he had worked as a research aide at Arizona State University and as a research intern at IIT, Guwahati. He was a systems engineer with the TCS Research and Innovation Labs Mumbai, until September 2020. He is currently working towards the Ph.D. degree at Department of Electrical Engineering, IIT Delhi. He is also a recipient of Prime Minister’s Research Fellows (PMRF) scheme, Ministry of Education, Government of India. His research interests include biomedical signal processing, human computer interface and machine learning.
\end{IEEEbiography}

\begin{IEEEbiography}
[{\includegraphics[width=1in,height=1.25in,clip,keepaspectratio]{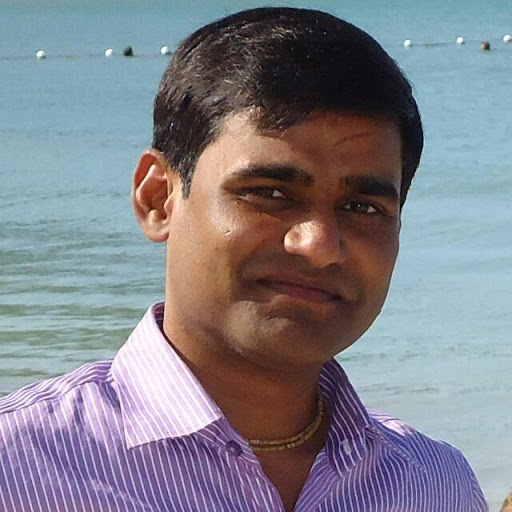}}]
{Lalan Kumar} (Member, IEEE)
is an Assistant Professor in the Department of Electrical Engineering, IIT Delhi, India. Prior to joining IIT Delhi, he has worked in IIT Bhubaneshwar as an Assistant Professor and a research fellow at Nanyang Technological University, Singapore. He has also worked at Motorola Bangalore as a software engineer with a multimedia team between 2008–2009. He received his B.Tech. degree in electronics engineering from Indian Institute of Technology (BHU), Varanasi, in 2008 and Ph.D. degree from Indian Institute of Technology, Kanpur, in 2015. His current area of research is Biomedical signal processing, Brain computer interface and microphone array processing for sound source localization. The detailed biographical information can be found at the URL: \url{http://web.iitd.ac.in/~lalank/}
\end{IEEEbiography}

\begin{IEEEbiography}
[{\includegraphics[width=1in,height=1.25in,clip,keepaspectratio]{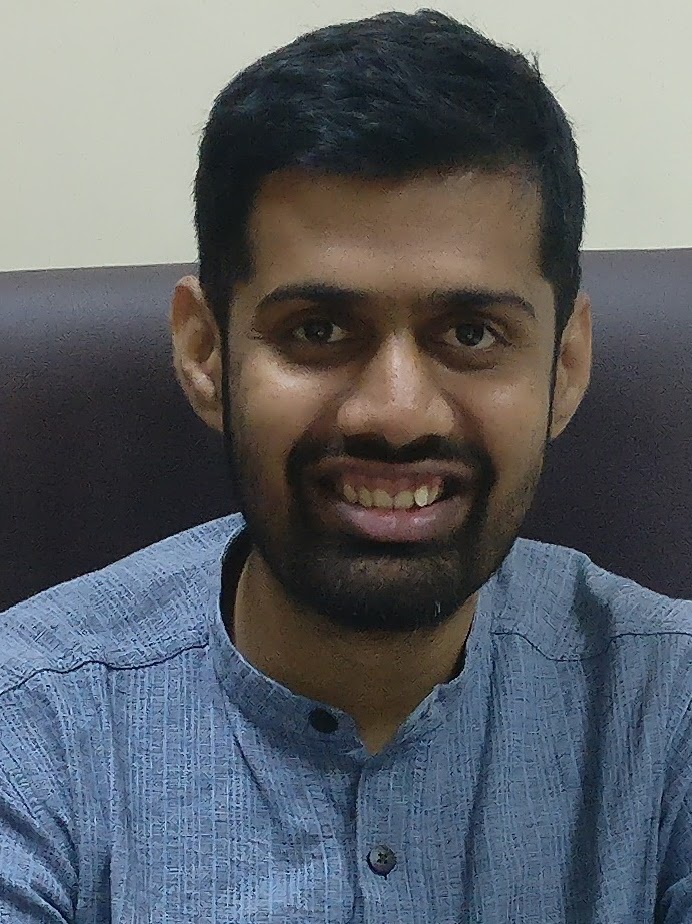}}]
{Prathosh A.P.} received his Ph. D from the Indian Institute of Science (IISc), Bangalore in 2015, in the area of temporal data analysis. He submitted his Ph. D thesis three years after his B.Tech in 2011, with many top-tier journal publications. Subsequently, he worked in corporate research labs including Xerox Research India, Philips research, and a start-up in CA, USA. His work in the industry, focussing on healthcare analytics, led to the generation of several IP, comprising 15 (US) patents of which 10 are granted and 6 are commercialized. He joined IIT Delhi in 2017 as an Assistant Professor in the computer technology group. Currently, he is a faculty member at the department of ECE at IISc, Bangalore. His current research includes deep-representational learning, cross-domain generalization, signal processing, and their applications in computer vision and speech analytics. 
\end{IEEEbiography}

\begin{IEEEbiography}
[{\includegraphics[width=1in,height=1.25in,clip,keepaspectratio]{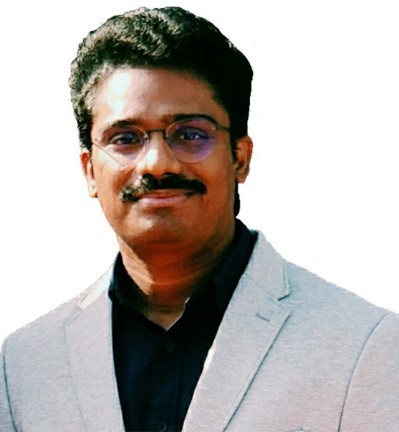}}]
{Suriya Prakash Muthukrishnan}  received his MD in Physiology from the All India Institute of Medical Sciences (AIIMS), New Delhi in 2014. He is now a Faculty of Physiology at AIIMS New Delhi. His research interests include investigating the neurophysiological mechanisms underlying visuospatial working memory and object recognition using functional and effective EEG connectivity of brain networks, identifying cost-effective biomarkers for diagnosis and prognosis of brain disorders, and devising novel treatment strategies using neuromodulation and neurofeedback in neurodegenerative and neuropsychiatric disorders.
\end{IEEEbiography}

\end{document}